\documentclass[twocolumn]{article}

\usepackage[T1]{fontenc} 
\usepackage{microtype} 

\usepackage[english]{babel} 

\usepackage[hmarginratio=1:1,top=32mm,columnsep=20pt,left=20mm]{geometry} 
\usepackage[]{caption} 
\usepackage{booktabs} 

\usepackage{enumitem} 
\setlist[itemize]{noitemsep} 

\usepackage{abstract} 

\usepackage{titlesec} 
\renewcommand\thesection{\Roman{section}} 
\renewcommand\thesubsection{\roman{subsection}} 
\titleformat{\section}[block]{\large\scshape\centering}{\thesection.}{1em}{} 
\titleformat{\subsection}[block]{\large}{\thesubsection.}{1em}{} 

\usepackage{titling} 
\usepackage{chemformula} 

\pretitle{\begin{center}\Large\bfseries} 
\posttitle{\end{center}} 
\title{Tunable nano-plasmonic photodetectors} 
\author{%
\textsc{Patrick Pertsch, René Kullock, Vinzenz Gabriel, Luka Zurak}\\
\textsc{Monika Emmerling and Bert Hecht}\\
[1ex] 
\normalsize NanoOptics \& Biophotonics Group, Experimental Physics 5,\\
\normalsize University of Würzburg, Am Hubland, 97074 Würzburg, Germany \\ 
\normalsize {hecht@physik.uni-wuerzburg.de} 
}
\date{\today} 
\renewcommand{\maketitlehookd}{%
\begin{abstract}
\noindent Visible and infrared photons can be detected with a broadband response via the internal photoeffect.
By using plasmonic nanostructures, i.e. nanoantennas, wavelength selectivity can be introduced to such detectors through geometry-dependent resonances. Also, additional functionality, like electronic responsivity switching and polarization detection have been realized.
However, previous devices consisted of large arrays of nanostructures to achieve detectable photocurrents.

Here we show that this concept can be scaled down to a single antenna level, resulting in detector dimensions well below the resonance wavelength of the device.
Our design consists of a single electrically-connected plasmonic nanoantenna covered with a wide-bandgap semiconductor allowing broadband photodetection in the VIS/NIR via injection of hot carriers. We demonstrate electrical switching of the color sensitivity as well as polarization detection. Our results hold promise for the realization of ultra small, highly integratable photodetectors with advanced functionality.
\end{abstract}
}

\begin{document}

\maketitle


\section{Introduction}
Active elements in modern integrated electronic circuits have reached sizes of only a few tens of nanometers \cite{noauthor_international_2021}, thanks to the short de Broglie wavelength of the conduction electrons.
Optical elements, on the other hand, are limited by diffraction effects, leading to much larger device sizes typically beyond the scale of the photon wavelength \cite{fossum_review_2014, kim_56_2020}.
Yet it is desirable to integrate electronic and photonic elements, since light is widely employed as signal carrier in modern telecommunication applications \cite{mitschke_fiber_2016}.
This in particular includes photodetectors, which convert optical into electrical signals.
However, shrinking photodetectors below the diffraction limit is not easily possible without loss in efficiency.
First of all, reducing the thickness of the detector leads to decreasing light absorption in its active material \cite{kim_56_2020}.
This especially applies for silicon, due to its indirect bandgap, but also to other materials, like few-layer transition-metal dichalcogenides, which are promising for modern photonic applications and narrow-band photodetectors, due to their comparatively strong absorption at their exciton transition energies \cite{wang_electronics_2012,mak_photonics_2016}.
Secondly, if the lateral size is reduced below the diffraction limit, the absorption in the device will decrease due to the mismatch between the detector cross section and the area of the incoming light field.
Finally, electrical connections have to be attached to the active material, which can lead to shadowing of the detection-area, further reducing the device efficiency \cite{hetterich_optimized_2007, lin_metal-insulator-semiconductor_2010}.

All of these problems can be circumvented by employing electrically-connected plasmonic nanoantennas as light collecting elements, since they provide significant antenna gain, i.e. they channel light from an effective cross section that is larger than their geometrical footprint \cite{novotny_principles_2012}.
This leads to more efficient utilization of the incoming light even for small lateral sizes and at low thicknesses.
In addition, the metallic antenna itself can be used as electrode, reducing the shadow effect imposed by electrical connections \cite{hetterich_optimized_2007}.
Also, plasmonic nanoantennas can be used directly as source of hot electrons and holes that are created inside the antenna material upon absorption of light \cite{narang_plasmonic_2016}.
By combining the antennas with a semiconducting material, these hot carriers will contribute to a photocurrent via the internal photoeffect (IPE).
This approach has been used for photodetectors with a broadband response in the visible (VIS) and near infrared (NIR) regime \cite{gao_porous_2019,tagliabue_hot-hole_2020,wang_hot_2015}, independent of the bandgap of the used semiconductor (SC), while the resonant behavior of plasmonic nanoantennas has allowed to selectively enhance the detection of a defined wavelength range, which can be tuned via the antenna geometry \cite{knight_photodetection_2011, hou_hot-electron_2021, li_circularly_2015}.
Moreover electronic responsivity switching \cite{wang_hot_2015} and polarization detection \cite{hou_hot-electron_2021, li_circularly_2015} have been reported for plasmonic IPE detectors.
However, to achieve reasonable photocurrents, large arrays of nanostructures had to be used so far \cite{huang_low-dimensional_2018}.

Here we show that the IPE in a single nanoantenna device can be utilized for photodetection at the diffraction limit by combining electrically connected gold nanoantennas with titaniumoxide ( \ch{TiO_x} ) as semiconducting material.
This material stack allows a broadband VIS/NIR response while the antenna geometry determines the specific absorption wavelength.
Further, we are able to electrically select this wavelength and determine the polarization angle of linearly polarized photons by employing different antenna geometries.
Our results are promising for ultra-small photodetectors with potential application, e.g., in high-resolution laser beam profiling.

\section{Results and discussion}
\begin{figure}
    \begin{center}
    \includegraphics[]{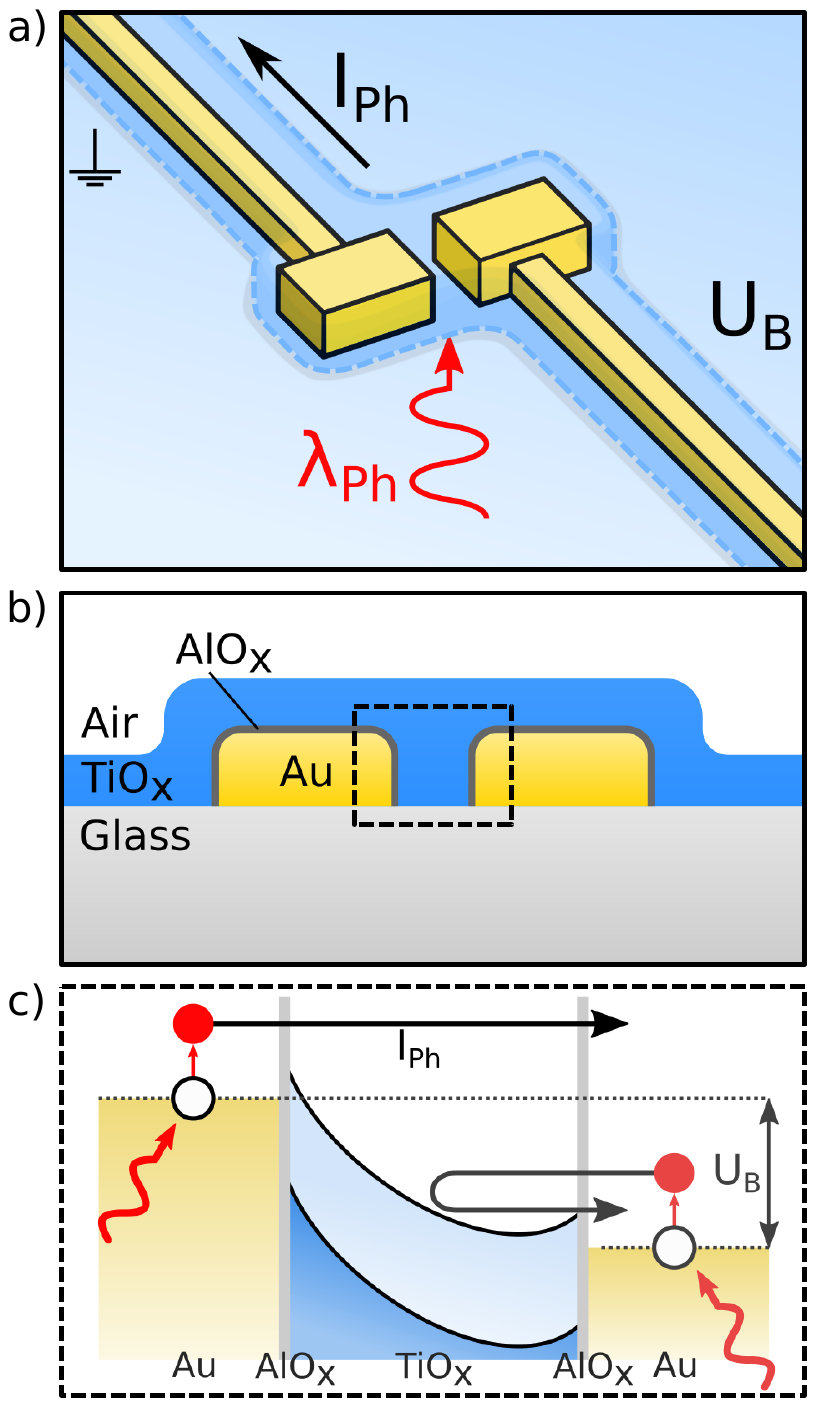}
    \end{center}

  \caption{Nanoantenna photodetector concept. a) Schematic representation of the device consisting of two \ch{Au} rods at the center, electrically connected via two wires and coated with \ch{TiO_x}. b) Cross section through the \ch{Au} rods perpendicular to the substrate surface. c) Energy-diagram of the metal-semiconductor-metal structure, under bias voltage $U_B$, with the excitation of hot carriers and the direction of the induced photocurrent $I_{Ph}$.}
  \label{fgr:schematics}
\end{figure}

In the simplest case, our device consist of a single dimer antenna on a glass substrate, like shown in figure \ref{fgr:schematics} a), electrically connected via two wires and coated with a semiconducting material to facilitate IPE.
We structure the antenna from a mono-crystalline gold flake \cite{krauss_controlled_2018} via focused ion beam milling and conformally coat it with \ch{TiO_x} in an atomic layer deposition process.
The thickness of the \ch{TiO_x} layer is chosen to fill the gap between the two arms of the antenna.

Since the direct growth of \ch{TiO_x} on gold structures led to ohmic contacts, we introduced a thin interfacial layer of aluminumoxide (\ch{AlO_x}) for interface passivation, which drastically lowers the dark current \cite{garcia_de_arquer_large-area_2015}.
With this additional layer we observe a Schottky-Barrier (SB) of $\sim 1.0$\,eV, determined from temperature dependent dark-current measurements.

Figure \ref{fgr:schematics} c) shows the energy band diagram of the device with applied bias voltage $U_B$. This metal-semiconductor-metal (MSM) structure features two SBs, one at each antenna arm, leading to low dark current at moderate bias (see SI).

Photocurrent measurements were performed with excitation-wavelengths between 500-900\,nm, which cannot directly lift electrons above the bandgap within the \ch{TiO_x} layer \cite{devlin_broadband_2016, pascual_fine_1978}.
However, hot carriers are excited within the metallic antenna arms and may be injected into the SC via the IPE \cite{zheng_distinguishing_2015}.
Due to the separation of holes and electrons at the SB, the electron flow is directed outwards, so from gold to the SC, while the electrons are subsequently collected by the other antenna arm.
Therefore, the photocurrent is a superposition of the contributions from both antenna-arms, with opposing current polarity, leading to a zero net-current for symmetric devices.
But this symmetry can be broken by applying a bias Voltage $U_B$, resulting in a detectable photocurrent.

\begin{figure*}[h]
    \begin{center}
    \includegraphics[]{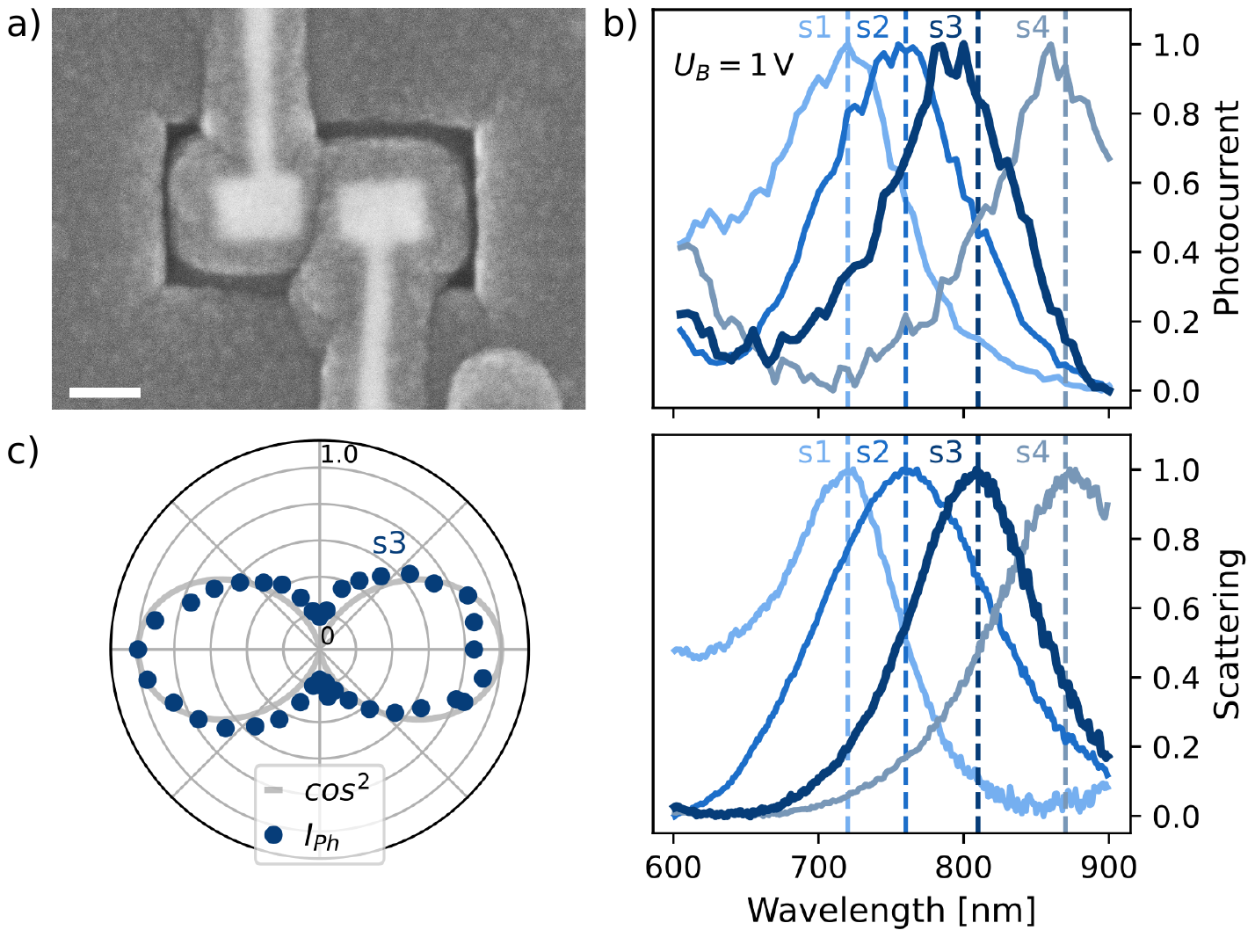}
    \end{center}

  \caption{Antenna enhanced photocurrent. a) SEM image of a fabricated nanoantenna covered with \ch{TiO_x}. The scale bar is 50\,nm. b) Comparison between the normalized photocurrent (top) and the scattering (bottom) of four antennas with different dimensions. c) Photocurrent as a function of the laser polarization angle at the resonance wavelength of 790\,nm. The antenna in a) is denoted as "s3" in the other plots.}
  \label{fgr:enhanced_photocurrent}
\end{figure*}

To proof the capability of this design to create very small photodetectors, we fabricated the symmetric dimer antenna displayed in the scanning electron micrograph (SEM) in figure \ref{fgr:enhanced_photocurrent} a) with a total length of 210\,nm including the \ch{TiO_x} layer. The size of the antenna is thus smaller than half of the shortest wavelength used during our measurements (500\,nm).

The bright areas in the image correspond to the gold of our antenna while the light gray areas represent the \ch{TiO_x} coating.
The dark rectangular frame around the antenna is a trench resulting from the ion-milling process and does not have a significant influence on our measurements.

The photocurrent for this antenna was measured at $U_B$=1\,V with tunable monochromatic light from a supercontinuum laser (see SI).
The resulting spectrum (s3 in figure \ref{fgr:enhanced_photocurrent} b) features a clear peak around 800\,nm that agrees well with the resonance wavelength in the scattering measurement.

By changing the height and length of the antenna, the resonance can be precisely controlled over a wide spectral range, which is demonstrated by the four scattering spectra, from different antennas, featuring distinct peaks between 600-900\,nm.
Longer resonance wavelengths were achieved by increasing the aspect ratio of the antenna arms.
The photocurrent peaks of all antennas coincide with the scattering resonances, showing that the color-selectivity of our detectors can be tuned via the antenna-dimensions.

The electrical connection wire also features a resonance in the scattering and photocurrent spectrum below 600\,nm.
This contribution is similar for all discussed antennas.
For clarity, only the resonances of the dimers are shown. 
A full photocurrent spectrum can be found the in the SI.

The peaks in the photocurrent spectrum can be explained by the increased absorption at the plasmonic resonance, which results in a high rate of hot carriers around this spectral position \cite{knight_photodetection_2011}.
This already shows that the antenna has a large influence on the properties of the detector.
As a further consequence, the polarization pattern of the detector is defined by the antenna geometry as well.
This is evident from the polarization dependent photoresponse depicted in figure \ref{fgr:enhanced_photocurrent} c), which can be approximately described by $\cos(\Phi)^2$, with $\Phi$ being the angle of polarization of the incident light relative to the antenna axis (x-direction in figure \ref{fgr:enhanced_photocurrent} a).
The obtained two-lobed pattern, recorded at the resonance wavelength of the antenna, is characteristic for a dipolar antenna \cite{novotny_principles_2012} and in agreement with previous reports \cite{knight_photodetection_2011}.
Note that the data was corrected for a decay of the photocurrent over time. Details on this can be found in the SI.

%
%
\begin{figure*}[!ht]
    \begin{center}
    \includegraphics[]{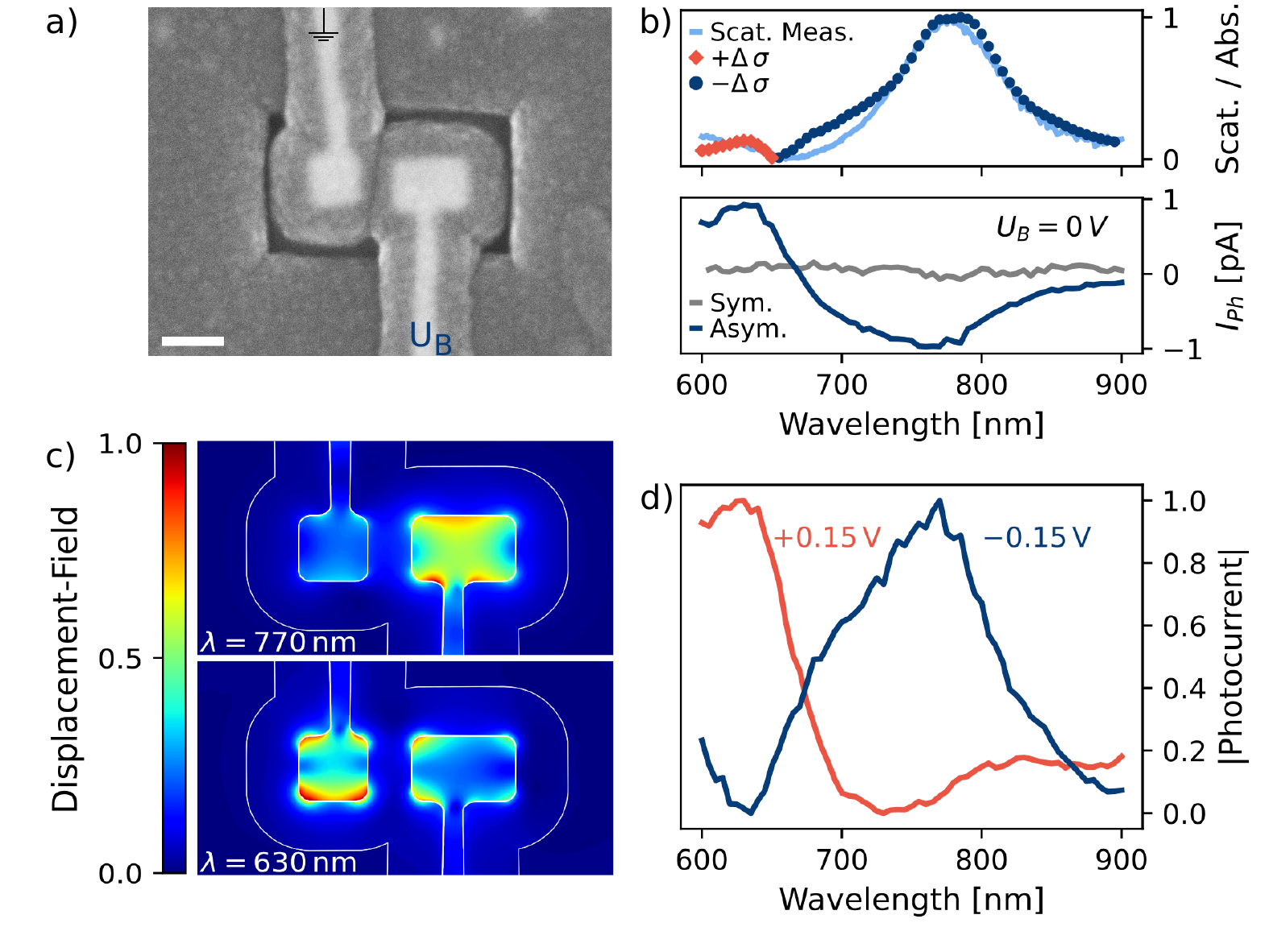}
    \end{center}

  \caption{Electrical responsivity switching. a) SEM image of an asymmetric antenna. The scale bar is 50\,nm. b) Measured scattering and simulated absorption spectra (top), plotted is the \emph{difference} of the absorption in both antenna arms $|\Delta\sigma|$, and photocurrent at zero bias (bottom). The latter is compared to the zero bias current of the symmetric antenna s3 from figure \ref{fgr:enhanced_photocurrent}. c) Simulated displacement fields (magnitude) of an asymmetric antenna at both resonances. The white lines denote the borders of the antenna and the \ch{TiO_x} layer. d) Electrical responsivity switching of the normalized photocurrent under positive and negative bias.}
  \label{fgr:switching}
\end{figure*}

Besides the static tuning of the detector via the geometry, our concept also allows a dynamic switching of the color-response by applying a voltage.
To this end, the asymmetric antenna in figure \ref{fgr:switching} a) is employed, which consist of two arms of different length, leading to two resonances in the device.
While the longer arm contribution is clearly visible in the scattering spectra in figure \ref{fgr:switching} b), the short arm resonance is hardly visible due to its small size and the thus reduced scattering efficiency.
We therefore performed FEM simulations, to prove the existence of the second resonance, taking the dimensions from the SEM image for our model.
As seen from the displacement field plots in figure \ref{fgr:switching} c), the results indeed show two modes, at 630\,nm and 770\,nm respectively, with one antenna arm excited predominantly in either case.

As mentioned before, the photocurrent of a \emph{symmetric} antenna is zero without applied bias voltage, which is visible in the gray curve in figure \ref{fgr:switching} b).
However, for the case of the \emph{asymmetric} dimer, both arms contribute currents with non-equal amplitude, depending on the excitation wavelength.
For example at 630\,nm, the short arm contributes more hot carriers as it is on resonance at this wavelength, while the long arm contributes more current at 770\,nm.
This leads to a non-zero net photocurrent with opposite polarity at the two resonances (see blue curve in the bottom plot of figure \ref{fgr:switching} b).

For comparison, we simulated the absorption in each arm separately, denoted as $\sigma_{left}$ and $\sigma_{right}$, which is proportional to the hot carrier generation inside the respective arm.
Therefore the difference $\Delta\sigma = \sigma_{left} - \sigma_{right}$ resembles the net-photocurrent at zero bias.
For comparability with the scattering, the absolute value $|\Delta\sigma|$ is added to the top plot in figure \ref{fgr:switching} b), while the sign is color encoded.
Although this simple model can predict the spectral positions of the peaks, it neglects important effects, like the wavelength dependent injection rate of hot carriers \cite{knight_photodetection_2011} or the scattering of hot carriers inside the metal \cite{narang_plasmonic_2016}.
To fully describe the photocurrent amplitude, a more extensive model is necessary, which is beyond the scope of this work.

To use the asymmetric antenna for dynamic color switching, a bias voltage $U_B$ is applied.
This induces a potential barrier, which suppresses the hot carriers from one of the two arms \cite{wang_power-independent_2013,wang_hot_2015}, as depicted in figure \ref{fgr:schematics} c), reducing its contribution to the net current.
This way, the carriers originating from the other antenna arm, on the higher electrical potential, are the main contributors to the photocurrent, and consequently, this arm defines the spectral response of the detector.
Therefore, by switching the sign of $U_B$, we can switch between the two resonances of the asymmetric antenna, which clearly changes the color response of the device (figure \ref{fgr:switching} d).

To completely suppress the photocurrent peak at a specific wavelength, the applied bias needs to be chosen carefully.
The simulations in \ref{fgr:switching} c) show that at, say, 770\,nm the excitation is spread over both arms, while one is clearly dominant.
So, while the main part of the light is absorbed in the long arm, a small fraction is still absorbed in the short arm.
Consequently, at sufficiently high bias, the short arm defines the photocurrent at 770\,nm, although it is not at resonance, because the long rod´s hot electrons are blocked by the potential barrier.
The optimal voltage, in our case $U_B=\pm 0.15\,V$, is therefore reached when the photocurrents of both arms cancel out at one resonance (the unwanted one). 
Choosing other values of $U_B$ can be used to tune the relative amplitude of the photocurrent peaks.

%
%
\begin{figure*}[!ht]
    \begin{center}
    \includegraphics[]{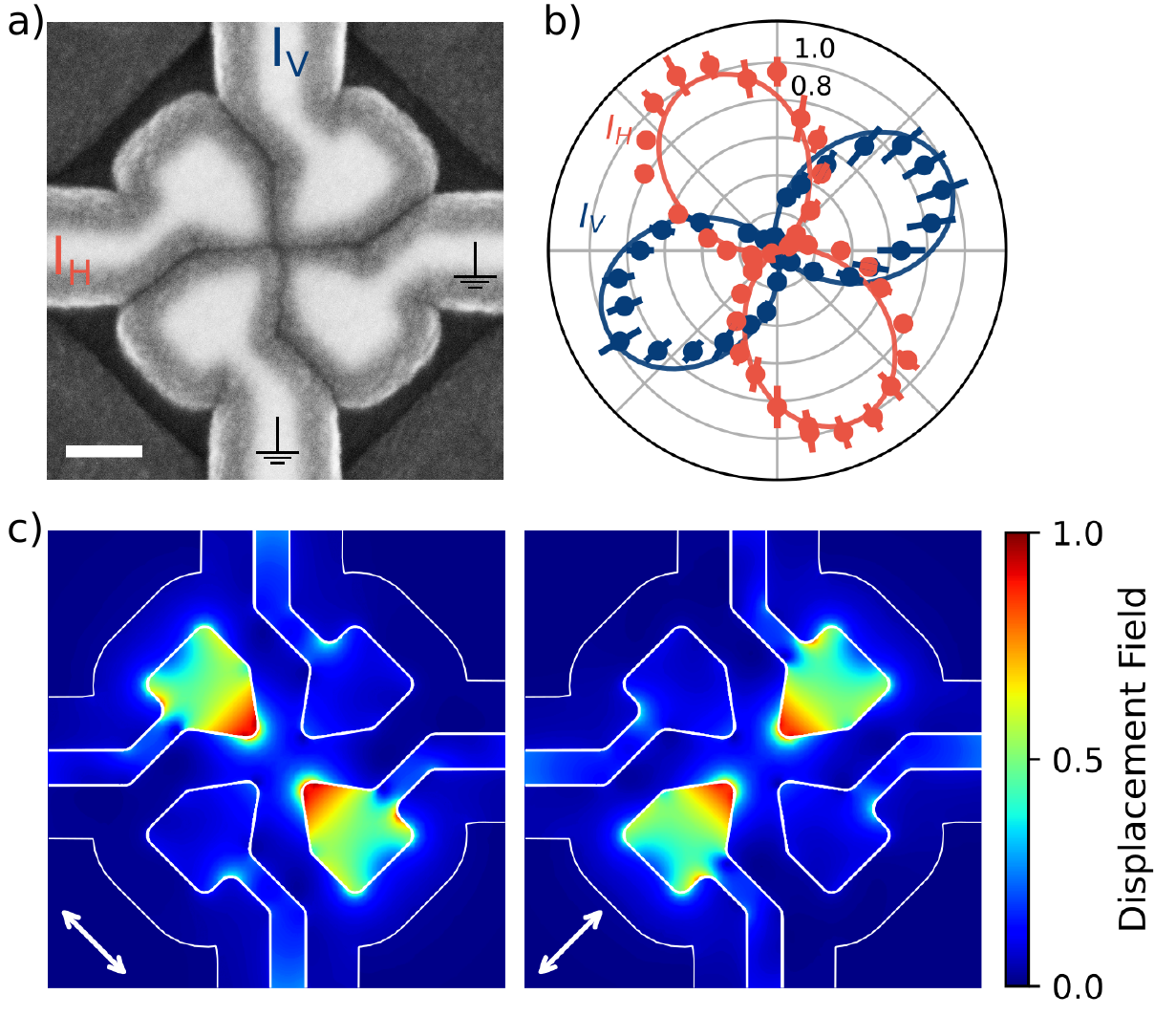}
    \end{center}

  \caption{Polarization detection. a) SEM image of a cross antenna and b) its photocurrent as a function of the laser polarization angle. The scale bar is 50\,nm. The currents $I_{H,V}$ through the horizontal and vertical connectors are measured separately and then normalized to the total current $I_H+I_V$. c) Simulated displacement fields, at the resonance wavelength of 800\,nm, for two orthogonal, linear polarizations, along the two dimers of the cross antenna.}
  \label{fgr:polarization}
\end{figure*}

We now show that our plasmonic photodetector concept also allows to detect the polarization angle $\Phi$ of linearly polarized light within a sub-diffractional device.
The polarization pattern in figure \ref{fgr:enhanced_photocurrent} c) already shows that the photocurrent of the dimer antenna depends on $\Phi$, so, in principle, this can be used as a simple polarization detector already.
If the power $P$ of the incoming light is constant and the angular dependence of the photocurrent $I_{Ph}(\Phi)$ is known, we can deduce the polarization angle $\Phi$ by measuring $I_{Ph}$.
However, for the simple case of the dimer, $I_{Ph}$ depends on $\Phi$ \emph{and} $P$. Thus, for variable $P$, the polarization angle cannot be deduced from $I_{Ph}$ alone.

To overcome this problem, the antenna design depicted in figure \ref{fgr:polarization} a) can be employed.
This cross antenna consists of two dimers that are rotated by 90° with respect to each other \cite{biagioni_near-field_2009}, while the current through the antenna arms can be measured separately via the connection wires.
The currents through the two dimers are denoted as $I_H$ and $I_V$ respectively (cf. figure \ref{fgr:polarization} a).
The resulting polarization patterns at the resonance wavelength are shown in figure \ref{fgr:polarization} b).
These two separate patterns can be described well by $I_V=\cos^2(\Phi+\Phi_0)$ and $I_H=\cos^2(\Phi+\Phi_0 + 90^{\circ}) = \sin^2(\Phi+\Phi_0)$ respectively, which indicates an independent excitation of both dimers.
The phase shift $\Phi_0$ results from the overall rotation of the cross.

This independence is supported by FEM simulations for the cross antenna:
The displacement-field plots in figure \ref{fgr:polarization} c) were computed for two orthogonal polarization directions of the incoming light.
In both cases the excitation of the dimer perpendicular to the polarization is negligible.

To measure $\Phi$ we now use both photocurrents $I_H$ and $I_V$, which again depend on the power of the incoming light and its polarization angle.
However, the total current through the cross $I_{tot}=I_V + I_H \propto cos(\Phi)^2 + sin(\Phi)^2 = 1$ changes with power but does not depend on the polarization angle.
Thus, by normalizing $I_{V,H}$ by $I_{tot}$, we arrive at a quantity that does depend on the polarization angle \emph{only}, assuming that the amplitude of both photocurrent-patterns is the same for both dimers.
Therefore, $I_{Ph}(\Phi,P)$ only needs to be measured once, at a constant light power, substantially reducing the necessary calibration effort for the cross antenna.
As an experimental proof, we measured the polarization pattern for different powers of the excitation laser (see SI) and observed no significant changes in the normalized photocurrent, while the absolute values increased with higher powers.
This shows that the cross antenna can be used as an intensity-invariant polarization detector.

Due to the four-fold symmetry, the cross antenna cannot uniquely resolve the angle of polarization.
But this problem can be overcome by using an antenna with three-fold symmetry, like shown in the SI.
The antenna presented there can resolve the polarization angle unambiguously, however at the cost of higher measurement complexity.

%
%
The antennas discussed in this work were fabricated to feature resonances in the range between 600-900\,nm, which was limited by our excitation source, towards long wavelengths, and by the increasing absorption of gold below 600\,nm.
But in principle this wavelength range can be easily extended to the infrared by further increasing the aspect ratio of the antennas \cite{novotny_principles_2012}, in theory up to the SB height\cite{knight_photodetection_2011}, which in our case is 1.0\,eV.
Plasmonic photodetectors at blue wavelengths are also feasible by employing different materials, like aluminum or silver \cite{knight_aluminum_2014}.

A bigger limitation, in terms of possible applications, is the detection efficiency which is rather low compared to standard photodetectors based on silicon. 
We calculate the detection efficiency by dividing the photocurrent by the total incoming power of our laser light, corrected by the size of the laser spot and the antenna absorption cross section. See SI for details.
With this we estimate an external quantum efficiency $\eta_{ext} \approx10^{-4}$ electrons per photon.

To a certain degree this efficiency is influenced by the thin \ch{AlO_x} layer, but since it is comparable to other reports \cite{knight_photodetection_2011}, the layer´s direct impact on $\eta_{ext}$ seems to be rather insignificant, due to the small thickness of only 0.5\,nm.
More importantly, the injection of hot carriers into the semiconductor is believed to be the biggest limitation for plasmonic IPE detectors \cite{narang_plasmonic_2016}, although covering the antenna with the SC should slightly increase injection probability, compared to placing it on a flat SC surface \cite{knight_embedding_2013}.

In the future, this injection might be improved by the so-called "plasmon-induced interfacial charge-transfer transition", which features internal quantum efficiency of $>24\%$ \cite{wu_efficient_2015, furube_ultrafast_2007, long_instantaneous_2014, tan_plasmonic_2017}.
Another approach to increase the detection efficiency could be via an integrated gain mechanism, like demonstrated for an IPE detector based on \ch{MoS2} with a reported photogain of $10^5$ \cite{wang_hot_2015}.

%
%
In conclusion, we showed, that plasmonic IPE devices can be scaled down to a single antenna level, enabling detectors with sizes well below the wavelength of the detected light.
In addition we demonstrated that these detectors can exhibit complex functionality like dynamic color switching and polarization detection, while keeping the small device footprint.
This renders our concept promising for small scale photodetectors, especially for applications where high resolution demands are combined with high light intensity, like, for example, the profiling of focused laser beams.
\bibliographystyle{plain}

\bibliography{pertsch2022-tunable}


\end{document}